\documentclass[final]{aipproc}

\layoutstyle{8x11single}

\begin{document}

\title{An Alternative View of the Dynamical Origin of the $P_{11}$ Nucleon Resonances: Results from the Excited Baryon Analysis Center}


\author{Hiroyuki Kamano}{
address={Department of Physics, Osaka City University, Osaka 558-8585, Japan},
altaddress={Excited Baryon Analysis Center, Thomas Jefferson National Accelerator Facility, Newport News, Virginia 23606, USA} 
}

\begin{abstract}
We present an alternative interpretation for the dynamical origin of the $P_{11}$ nucleon 
resonances, which results from the dynamical coupled-channels analysis at Excited Baryon 
Analysis Center of Jefferson Lab.
The results indicate the crucial role of the multichannel reaction dynamics in determining
the $N^\ast$ spectrum.
\end{abstract}

\maketitle

\paragraph{Introduction}
An understanding of the spectrum and structure of the excited nucleon ($N^*$) states is 
a fundamental challenge in the hadron physics. 
The $N^*$ states, however, couple strongly to the meson-baryon continuum states and 
appear only as resonance states in the $\gamma N$ and $\pi N$ reactions. 
One can expect from such strong couplings that the (multichannel) reaction dynamics 
will affect significantly the $N^\ast$ states and
cannot be neglected in extracting the $N^*$ parameters from the data and 
giving physical interpretations. 
It is thus well recognized nowadays that a comprehensive study of 
all relevant meson production reactions with $\pi N, \eta N, \pi\pi N, KY, \cdots$ 
final states is necessary for a reliable extraction of the $N^*$ parameters.

To address this challenging issue,
the Excited Baryon Analysis Center (EBAC) of Jefferson Lab 
has been conducting the comprehensive analysis of the world data of 
$\gamma N,\pi N\to \pi N, \eta N, \pi\pi N, KY, \cdots$ reactions
systematically, covering the wide energy and kinematic regions. 
The analysis is pursued with a dynamical coupled-channels (DCC) model, 
the EBAC-DCC model~\cite{msl07}, 
within which the unitarity among relevant meson-baryon channels, 
including the three-body $\pi\pi N$ channel, is fully taken into account.

Within the EBAC-DCC model, the partial wave amplitudes of
$M(\vec p) + B(-\vec p) \to M'(\vec p') + B'(-\vec p')$ 
with $MB,M'B'= \pi N,\eta N,\pi\Delta,\sigma N,\rho N,\cdots $
are calculated by the following coupled-channels integral equations
(suppressing the angular momentum and isospin indices):
\begin{equation}
T_{MB,M'B'}(p,p';E) = V_{MB,M'B'}(p,p';E)
+\sum_{M''B''}\int dq q^2 V_{MB,M''B''}(p,q;E) G_{M''B''}(q;E) T_{M''B'',M'B'}(q,p';E).
\label{eq:lseq}
\end{equation}
Here $E$ is the total energy in the center of mass frame;
$G_{MB}(q;E)$ is the Green function of the $MB$ channel,
which is expressed as 
$G_{MB}(q;E)=1/[E-E_M(q)-E_B(q) + i\epsilon]$
for the stable channels and
$G_{MB}(q;E)=1/[E-E_M(q)-E_B(q) - \Sigma_{MB}(q;E)]$
for the unstable $\pi\Delta$, $\rho N$, and $\sigma N$ channels.
The imaginary part of the self-energy, ${\rm Im}[\Sigma_{MB}(q,E)]$,
is determined by the three-body $\pi\pi N$ unitarity cut.
The $MB\to M'B'$ transition potential is defined by
\begin{equation}
V_{MB,M'B'}(p,p';E) = v_{MB,M'B'}(p,p') + 
\sum_{N^\ast_i} \frac{\Gamma^\dag_{N^\ast_i,MB}(p)\Gamma_{N^\ast_i,M'B'}(p')}{E-m^0_{N^\ast_i}},
\label{eq:pot}
\end{equation}
where $m^0_{N^\ast_i}$ and $\Gamma^\dag_{N^\ast_i,MB}(p)$
represent the mass and $N^\ast_i \to MB$ decay vertex
of the $i$-th bare $N^\ast$ state in a given partial wave, respectively.
The first term $v_{MB,M'B'}(p,p')$ is 
the meson-exchange potential,
which is derived from the effective Lagrangian by
making use of the unitary transformation method~\cite{msl07};
the second term describes $MB\to M'B'$ transitions 
through the bare $N^\ast$ state, $MB\to N^\ast\to M'B'$.

The $MB\to M'B'$ amplitude (\ref{eq:lseq}) is a basic ingredient to construct
all single and double meson production reactions
with the initial $\pi N$, $\gamma N$, $N(e,e')$ states.
The parameters of our model have been fixed by analyzing the $\pi N$ scattering~\cite{jlms07} 
up to $E=2$ GeV and $\gamma p\to\pi N$~\cite{jlmss08} and
$ep\to e'\pi N$~\cite{jklmss09} up to $E=1.6$ GeV, respectively. 
Then the model has been applied to $\pi N\to \pi\pi N$~\cite{kjlms09} and 
$\gamma N\to\pi\pi N$~\cite{kjlms09-2} to examine the consistency of 
our coupled-channels framework.

The pole positions of the resonance states can be obtained as
zeros of the determinant of the inverse of 
the dressed $N^\ast$ propagator:
\begin{equation}
[D^{-1}(E)]_{i,j} = (E - m^0_{N^*_i})\delta_{i,j} - [M(E)]_{i,j},
\label{eq:nstar-selfe}
\end{equation}
where the self-energy of the dressed $N^\ast$ propagator is given by
\begin{equation}
[M(E)]_{i,j}=
\sum_{MB}
\int q^2 dq 
\bar{\Gamma}_{N^*_j \to M B}(q;E) G_{MB}(q,E) {\Gamma}_{MB \to N^*_i}(q).
\equiv \sum_{MB}[M_{MB}(E)]_{i,j},
\label{eq:nstar-g}
\end{equation}
with $\bar\Gamma_{N^\ast\to MB}$ is the dressed $N^\ast\to MB$ vertex
defined in Ref.~\cite{msl07}.

Here it is noted that within our framework the bare $N^*$ states are defined as 
the eigenstates of the Hamiltonian in which the couplings to 
the meson-baryon continuum states are turned off~\cite{msl07}.
Therefore, by definition, our bare $N^*$ states can be related with the hadron states 
obtained from the static hadron structure calculations such as quark models.
Through the reaction processes, the bare $N^*$ states couple to the continuum states, 
and then they become resonance states.
Of course there is another possibility that the meson exchange processes arising from 
the first term of Eq.~(\ref{eq:pot}), $v_{MB,M'B'}(p',p)$, generate resonance poles dynamically.
Our framework contains both possibilities.
In general, the physical resonances will be a mixture of the two possibilities.

The $N^\ast$ pole positions extracted from the EBAC-DCC analysis in 2006-2010 
have been reported in Ref.~\cite{sjklms10}.
In the following, we focus on presenting our findings for the $P_{11}$ nucleon resonances.

\begin{figure}
\includegraphics[height=.25\textheight]{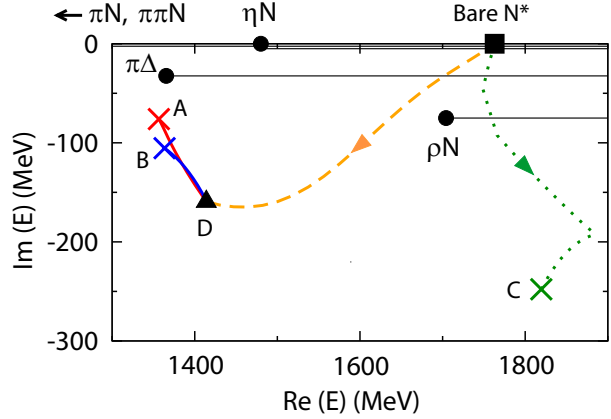}
\caption{Resonance poles in the $P_{11}$ partial wave below ${\rm Re}(E)=2$ GeV:
$1357-i76$ MeV (Pole A), $1364-i105$ MeV (Pole B), and $1820-i248$ MeV (Pole C). 
See the text for the description of the figure.}
\label{fig:trajectory}
\end{figure}

\paragraph{Dynamical origin of $P_{11}$ nucleon resonances}
As shown in Fig.~\ref{fig:trajectory}, we have found total three $P_{11}$ resonance poles 
in the complex-$E$ plane below ${\rm Re}(E)=2$ GeV.
An interesting result is that two resonance poles appear in the 1st $P_{11}$ resonance region
with ${\rm Re}(E) \sim 1350$ MeV, which are rather stable against large variations
of model parameters within our approach~\cite{knls10}.
The 1st $P_{11}$ resonance is famous as the Roper resonance, which is known to be 
one of the controversial baryon resonances.
Our result suggests that the Roper resonance is associated with the two resonance poles.
It is worthwhile to mention that other groups have also reported such 
a two-pole structure of the Roper resonance~\cite{vpi85,said06,cmb90,julich09}.

Another finding is that the two Roper poles (poles A and B in Fig.~\ref{fig:trajectory})
and the next higher resonance pole corresponding to
$N^*(1710)$ (pole C), are generated from a single bare state as a result of its coupling to 
the multichannel meson-baryon continuum states.
Theoretically, Eden and Taylor already pointed out more than four decades ago that 
the multichannel reaction dynamics can generate many resonance poles 
from a single bare state~\cite{eden}.
In most cases, only one of such poles appears close to the physical real energy axis.
However, depending on a reaction dynamics, more than one pole can appear 
to have a physical significance.
Just a few such evidences were reported in the past~(see, e.g., Ref.~\cite{morgan}).
Our result suggests that the $P_{11}$ resonances may be an important addition to it.

To examine how the three $P_{11}$ poles evolve from a single bare state 
dynamically, we trace the zeros of 
$\det[\hat D^{-1}(E)]=\det[E-m_{N^\ast}^0-\sum_{MB}y_{MB}M_{MB}(E)]$
in the region $0\leq y_{MB}\leq 1$,
where $M_{MB}(E)$ is the $MB$-loop contribution to
the $N^\ast$ self-energy $M(E)$ defined in Eq.~(\ref{eq:nstar-g}).
Each $y_{MB}$ is varied independently to find continuous evolution
paths through the various Riemann sheet on which the analytic 
continuation method is valid.

By setting all $y_{MB}$'s to slightly positive from zero, 
the bare state (the filled square at $E=1763$ MeV in Fig.~\ref{fig:trajectory}) 
couples to all $MB$ channels and many poles are generated
according to the discussion by Eden and Taylor~\cite{eden}.
One of them appears on the $\eta N$-unphysical, 
$\rho N$-unphysical, and $\pi\Delta$-unphysical sheet
and it moves to the pole C 
by further varying all $y_{MB}$ to one 
(the dotted curve in Fig.~\ref{fig:trajectory}).

Similarly we can trace how the two Roper poles evolve from the
same bare state.
It is instructive to see this by first keeping $y_{\pi\Delta}$ zero
and varying the other $y_{MB}$'s from zero to one,
which means that the coupling to the $\pi\Delta$ channel is turned off during the variation.
With this variation, we can trace another pole trajectory moving on
the $\eta N$-physical and $\rho N$-physical sheet 
along the dashed curve in Fig.~\ref{fig:trajectory}
from the bare position to the point D with Re$(E_D)\sim 1400$ MeV
(the filled triangle in Fig.~\ref{fig:trajectory}).
It is noted that the poles on the $\eta N$-physical sheet 
is far from the physical real energy axis above the $\eta N$ threshold,
while it is the nearest below the threshold.
Therefore this pole on the $\eta N$-physical sheet, 
moving from the bare position to the point D along the dashed curve,
becomes the resonance close to the physical real energy axis 
as a result of crossing the $\eta N$ threshold.
By further varying $y_{\pi\Delta}:0\to 1$, the trajectory splits into 
two trajectories: One moves to the pole A on the $\pi\Delta$ unphysical 
sheet and the other to the pole B on the $\pi\Delta$ physical sheet.
This indicates that the coupling to the $\pi\Delta$ channel is 
essential for the two-pole structure of the Roper resonance.
In this way, we observe that all the three $P_{11}$ resonance poles 
are connected to the same bare $N^\ast$ state at $E=1763$ MeV.

Comparing the values between the bare $N^\ast$ mass ($E = 1763$ MeV) and the Roper pole masses
[${\rm Re}(E)\sim 1350$ MeV], we observe that 
the reaction dynamics can produce a sizable mass shift. 
It often comes to an issue that the Roper mass appears very high in the static hadron structure calculations.
In our point of view, however, it is not so surprising because the reaction dynamics 
is not taken into account in those static calculations.

\paragraph{Summary}
We have presented an alternative view of the dynamical origin of 
the $P_{11}$ nucleon resonances, which is summarized as the following two findings:
The two-pole pole nature of the Roper resonance and the one-to-multi correspondence between the
bare $N^\ast$ state and the physical resonance states.
All of these findings are the consequence of the nontrivial multichannel reaction dynamics and 
thus indicate that we cannot neglect reaction dynamics in understanding the hadron spectrum.

\begin{theacknowledgments}
The author would like to thank 
B.~Juli\'a-D\'iaz, T.-S.~H.~Lee, A.~Matsuyama, S.~X.~Nakamura, T.~Sato, and N.~Suzuki
for their collaboration at EBAC.
This work is supported by the U.S. Department of Energy, Office of Nuclear Physics Division
under Contract No. DE-AC05-06OR23177, under which Jefferson Science Associates operates the Jefferson Lab.
This research used resources of the National Energy Research Scientific Computing Center, 
which is supported by the Office of Science of the U.S. Department of Energy 
under Contract No. DE-AC02-05CH11231, and resources provided on ``Fusion,'' 
a 320-node computing cluster operated by the Laboratory Computing Resource Center
at Argonne National Laboratory.
\end{theacknowledgments}

\bibliographystyle{aipproc}   

\end{document}